\documentclass[fleqn,10pt,twocolumn]{wlscirep}
\usepackage{times}
\usepackage{soul}
\usepackage{url}
\PassOptionsToPackage{hyphens}{url}\usepackage{hyperref}
\usepackage[utf8]{inputenc}
\usepackage{graphicx}
\usepackage{amsmath}
\usepackage{booktabs}
\usepackage{multirow}
\usepackage{xcolor}
\usepackage[switch]{lineno}
\usepackage{tikz}
\usepackage{helvet}
\usepackage{mcite}

\newcommand*{\myfont}{\fontfamily{phv}\selectfont}

\title{Neural network based order parameter for phase transitions {\color{black}and its applications in high-entropy alloys}}

\author[1,*]{Junqi Yin}
\author[1,2,*]{Zongrui Pei}
\author[2]{Michael C. Gao}

\affil[1]{Oak Ridge National Laboratory, Oak Ridge, TN37831, USA}
\affil[2]{National Energy Technology Laboratory, Albany, OR97321, USA}

\affil[*]{corresponding authors (equal contributions). yinj@ornl.gov, peizongrui@gmail.com}

\begin{abstract}

Phase transition is one of the most important phenomena in nature and plays a central role in materials design. All phase transitions are characterized by suitable order parameters, including the order-disorder phase transition. However, finding a representative order parameter for complex systems is nontrivial, such as for high-entropy alloys. Given variational autoencoder's (VAE) strength of reducing high dimensional data into few principal components, here we 
coin a new concept of ``VAE order parameter". We propose that the Manhattan distance in the VAE latent space can serve as a generic order parameter for order-disorder phase transitions. The physical properties of the order parameter are quantitatively interpreted and demonstrated by multiple refractory high-entropy alloys. Assisted by it, a generally applicable alloy design concept is proposed by mimicking the nature mixing of elements. Our physically interpretable ``VAE order parameter" lays the foundation for the understanding of and alloy design by chemical ordering. 

\end{abstract}

\begin{document}
\flushbottom
\maketitle

\thispagestyle{empty}




\section*{Introduction}

Since Gibbs laid the foundation for the theory of phase diagram in 1879 \cite{gibbs1879equilibrium}, material scientists have had the ability to design materials with required phases for targeted material properties, guided by phase diagram.
Among the most important phase transitions is the ordering transition of phases where only the translational symmetry is broken while the lattice type does not change. Understanding the ordering of atoms in solids has been a central topic of fundamental importance in materials science and physics \cite{Thomas1951,sourmail2005near,yang2020ultrahigh,chen2020unprecedented}, while the rise of high-entropy alloys (HEAs) \cite{Yeh2004,Cantor2004, ZHANGYong20141,george2019high,pei2018overview,pei2020statistics,hea-gpc} makes the topic even more widely acknowledged \cite{zhang2020short,yin2020yield, pei2020statistics,Koermann2016}. HEAs are alloys consist of four or more components in equal or near equal molar.
For one phase in HEA, although the number of the degrees of freedom is often huge,
the physics is still governed mainly by few principal variables that reflect the underlying symmetry of the system. {\color{black} Introduced by Landau in 1937  \cite{landau}}, order parameters are such principal variables that describe the transition point (a symmetry breaking) between ordered (less symmetric) and disordered (more symmetric) phases. An order parameter (i) should have distinct values for different phases, e.g., non-zero values for ordered phases and zero for disordered ones; and (ii) its second-order moment (susceptibility),  {\color{black} if it has a definition}, should {\color{black}show a jump} at the transition points. {\color{black} Traditionally, the ordering degree of a {\color{black}HEA} system is described by multiple Warren-Cowley order parameters \cite{OWEN2016155}, and each of them can reflect only part of the information and thus is not representative. It also suffers from the distance cutoff and the information for long-range order (LRO) is missing. Although this can be supplemented by a number of LRO parameters, a single order parameter that can generally capture the overall randomness degree is appealing, which is simpler to understand and more convenient for applications in alloy design.} The challenge is that defining such a single (not multiple) order parameter is often a nontrivial task, particularly for complex concentrated alloys with extremely large dimensions of configurational spaces, and remains an active topic of research for metallic materials \cite{PhysRev.77.669,OWEN2016155}.   
Machine learning provides new opportunities to the challenge. Based on neural networks, variational autoencoder (VAE) \cite{vae,vae-tutorial} is widely acknowledged to be capable of mapping high dimensional data into a latent space with few variables. The principal information of the data is encoded into those latent variables in the sense that the data can be reconstructed during the decoding process. Other techniques such as principal component analysis (PCA) have been used \cite{phase-detection} for the same purpose, but VAE is a more general and scalable approach because it can encode/decode the non-linearity in the data using neural networks. 
Therefore, VAE has been widely utilized in various domain applications \cite{deepdrivemd,*vae-anomaly,*vae-ising-nature,*related1, *related2} as a powerful unsupervised technique for cluster analysis, dimensionality reduction, anomaly detection, etc. However, variables in VAE latent space are often treated qualitatively, and few attention has been paid to the physical interpretation of those latent variables. There are early works \cite{vae-ising-pre, *related3, *related4} in connecting the order parameter with the VAE latent variables, however, the scope is limited to toy systems {\color{black}and qualitative descriptions}.
\begin{figure*}[!ht]
  \includegraphics[width=\linewidth]{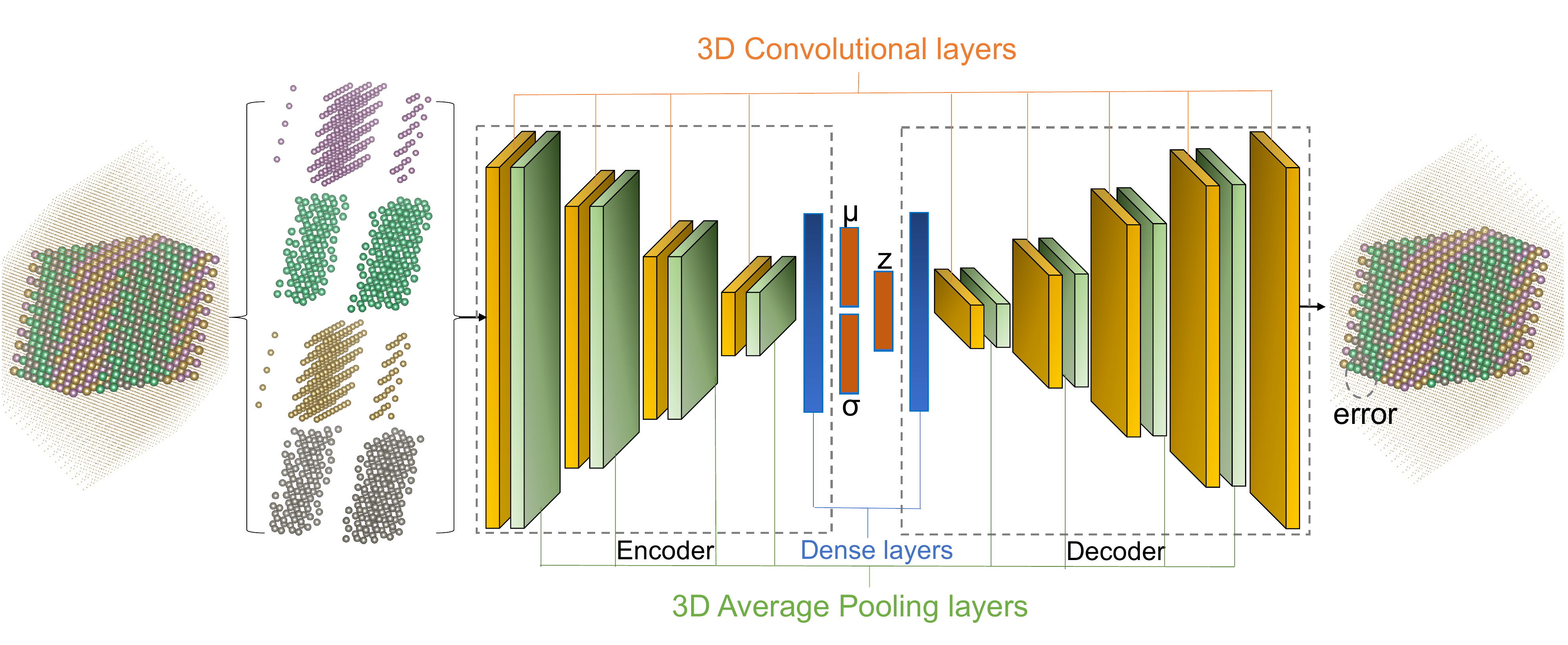}
  \caption{Neural network architecture of the VAE model. Both the encoder and decoder consist of pairs of a 3D convolutional layer followed by an average pooling layer except the last convolutional layer for output. The latent variable $z$ is sampled from a Gaussian distribution $\mathcal{N}(\mu, \sigma)$. The input is a cubic lattice (marked with small dots) with atoms sitting on BCC lattice sites (marked with filled circle). Each atom type is put into a separate channel. The illustration is for a four-component system with 4 channels  corresponding to Mo, Nb, Ta, and W (from top to down). The output is a reconstructed sample from the latent variable $z$ with only a couple of sites miss-classified.} 
  \label{fig:nn-arch}
\end{figure*}
We generalize the relationship between order parameters in phase transition and latent variables in VAE, and propose to use the Manhattan distance \cite{distance-metric} in latent space to characterize the order/disorder in alloys. We demonstrate the approach on representative body-centered cubic (bcc) HEAs with either four or five components and a novel bcc HEA proposed recently by nature mixing \cite{wei2020natural}. Refractory HEAs have attracted a lot of attentions \cite{wei2020natural,wang2020multiplicity,lee2020lattice,lee2020temperature,MoNbTaW-MRL,MoNbTaW-npjcm} because they can provide excellent mechanical properties. Since HEAs tend to be more mechanically favorable in disordered state, one of the challenges is to search for HEAs with lower order-disorder transition temperatures. To this end, reliable order parameters are needed to capture the signals of phase transitions, where our newly proposed VAE-based order parameter can offer great opportunities.


\section*{Results}

{\bf VAE-based new order parameter}

Defining a good metric for the ordering degree in an alloy system is non-trivial, and is an active research topic. The Warren-Cowley short-range order (SRO) parameter \cite{OWEN2016155} $\alpha_l^{ij}$ is currently the state of the art, which is defined as
$\alpha_l^{ij} = 1 - \frac{P^{i|j}_l}{c_i}$, 
where $c_i$ is the concentration of atom $i$, and $P^{i|j}_l$ is the probability of finding atom $i$ at the $l$-th neighbor for a given atom $j$. This quantity can be calculated for a given configuration and taken as an ensemble average. 
The shortcomings associated with Warren-Cowley order parameter are: 1) it is defined with respect to atomic pairs and a complete set of the order parameters is in the form of a matrix instead of a scalar value. In contrast to a matrix, a scalar number offers a more straightforward impression of the randomness of a system and is more acceptable; 2) it is parameterized over the neighboring distance, and it is unknown how many (or how long range) of the order parameters one should consider for a complete understanding of the ordering. The proper choice of it depends on the system.  




We propose a VAE-based order parameter to overcome these drawbacks. The structure of our VAE model is shown in Fig.~\ref{fig:nn-arch}. It is built up on 3D convolutional neural networks (CNN), consisting of encoder and decoder with the atom positions of HEAs as input. Following each feature-extraction CNN layer, a pooling layer is added to capture some translation invariance and prevent overfitting. Considering the sparsity of input data, the average operation is employed for the pooling layer instead of maximum, which performs better for sparse data \cite{Vasudevan:2018jd}. Additionally, we have tuned a few hyperparameters such as the number of CNN/pooling layers, kernel and stride size per layer, etc. Specific model parameters are listed in the supplementary materials. Our results indicate that pooling layers help to prevent the explosion of parameters for large alloy systems without much loss of accuracy.  

The atoms sit on a BCC lattice and the distances among them determine the strengths of interactions. Special attention is needed to ensure that CNN captures the correct features of lattice symmetry and distance.  To this end, the Monte Carlo (MC) generated configurations are transformed into a simple cubic lattice (see the input/output in Fig.~\ref{fig:nn-arch}) with sites being marked ``1" and ``0" depending on atomic occupations. In this way, the convolution operation counts the right neighbors and the physics is preserved. The relationship between the BCC unit vector ($a_1, a_2, a_3$) and simple cubic unit vector ($x_1, x_2, x_3$) is \cite{bcc}
\begin{linenomath}\begin{align}
    a_1 &= \frac{1}{2}(x_2 + x_3 - x_1); \notag \\
    a_2 &= \frac{1}{2}(x_3 + x_1 - x_2);  \\
    a_3 &= \frac{1}{2}(x_1 + x_2 - x_3),  \notag
\end{align}\end{linenomath}
hence the shape of the occupied sites is a rhombohedron. After this transformation, one BCC lattice is transformed into four lattices of simple cubic as the input $X$ for the VAE model. The input $X$ needs to contain both ordered and disordered configurations to train a generally applicable model.

Given the aforementioned features of this physical problem, the loss function of our VAE model is defined as 
\begin{linenomath}\begin{align}
\mathcal{L} = \mathcal{L}_{\mathrm{reconstruction}} + \mathcal{L}_{\mathrm{regularization}}, \label{loss-terms}
\end{align}\end{linenomath}
where the first term is the reconstruction loss between the input $X$ and model output $d(z)$ and the second one is a regularization term to prevent model from overfitting (See detailed discussion in the Method section). The two loss terms (Eq.~\ref{loss-terms}) behave differently for ordered and disordered phases. The reconstruction loss is less sensitive to randomized features ($d(z) \sim \textit{Constant}$ for random noise) comparing to ordered ones with distinct features, so the overall loss for disordered phases is dominated by the regularization term. For the Kullback-Leibler divergence (Eq.~\ref{eq:kl}) employed here as the regularizer, it follows 
\begin{linenomath}\begin{align}\label{eq:reg}
\mathcal{L} &\sim \mathcal{L}_{\mathrm{regularization}} \notag \\ 
                  &\sim \mu^2 - \log \sigma^2 + \sigma^2, 
\end{align}\end{linenomath}
the minimization of which leads to $\mu \Rightarrow 0$ for a given $\sigma$. In other words, if we hypothesize the symmetry of input $X$ is preserved by the latent variable $z$, the $D_{KL}$ loss is minimized when the more symmetric phases are encoded into data points near the origin of the latent space.  

Based on this observation, we propose to use the distance metric $(\sum_i|z_i|^p)^{1/p}$ in the latent space of a VAE as the order parameter for order-disorder phase transition. Considering lower $p$ is more preferable for high dimensional applications \cite{distance-metric}, the Manhattan distance (L1 norm) becomes a natural choice. The VAE based order parameter $Z^{op}$ at temperature $T$ is then defined as
\begin{linenomath}\begin{equation}
    \langle Z^{op}\rangle_T = \frac{1}{M}\sum_{X_j \in T}^M \sum_i^d|z_i(X_j)|, \label{eq:z}
\end{equation}\end{linenomath}
where $M$ is the number of samples and $d$ is the dimension of the latent space. {\color{black} Note that the distance in the VAE latent space is not equivalent to the spatial distance in the input data. It is rather a measure of feature similarities.} 

An order parameter should have non-zero values in ordered phase and go to zero in disordered phase. Furthermore, the second-order moment (i.e. susceptibility) of the order parameter should peak at the phase transition point. Similar to classical susceptibility, we define the VAE-based one,
\begin{linenomath}\begin{equation}
    \chi(Z^{op}) = \bigg{(}\langle(Z^{op})^2\rangle_T - (\langle Z^{op}\rangle_T)^2 \bigg{)}\frac{N}{T}. \label{eq:x}
\end{equation}\end{linenomath}
The second-order moment of energy (i.e. specific heat, $C_v = \frac{dE}{dT}$) is often a good indicator for phase transition as well. As will be seen below, our new VAE-based order parameter meets all physical requirements for an order parameter.





\begin{figure*}[!ht]
\centering
  \includegraphics[width=\linewidth]{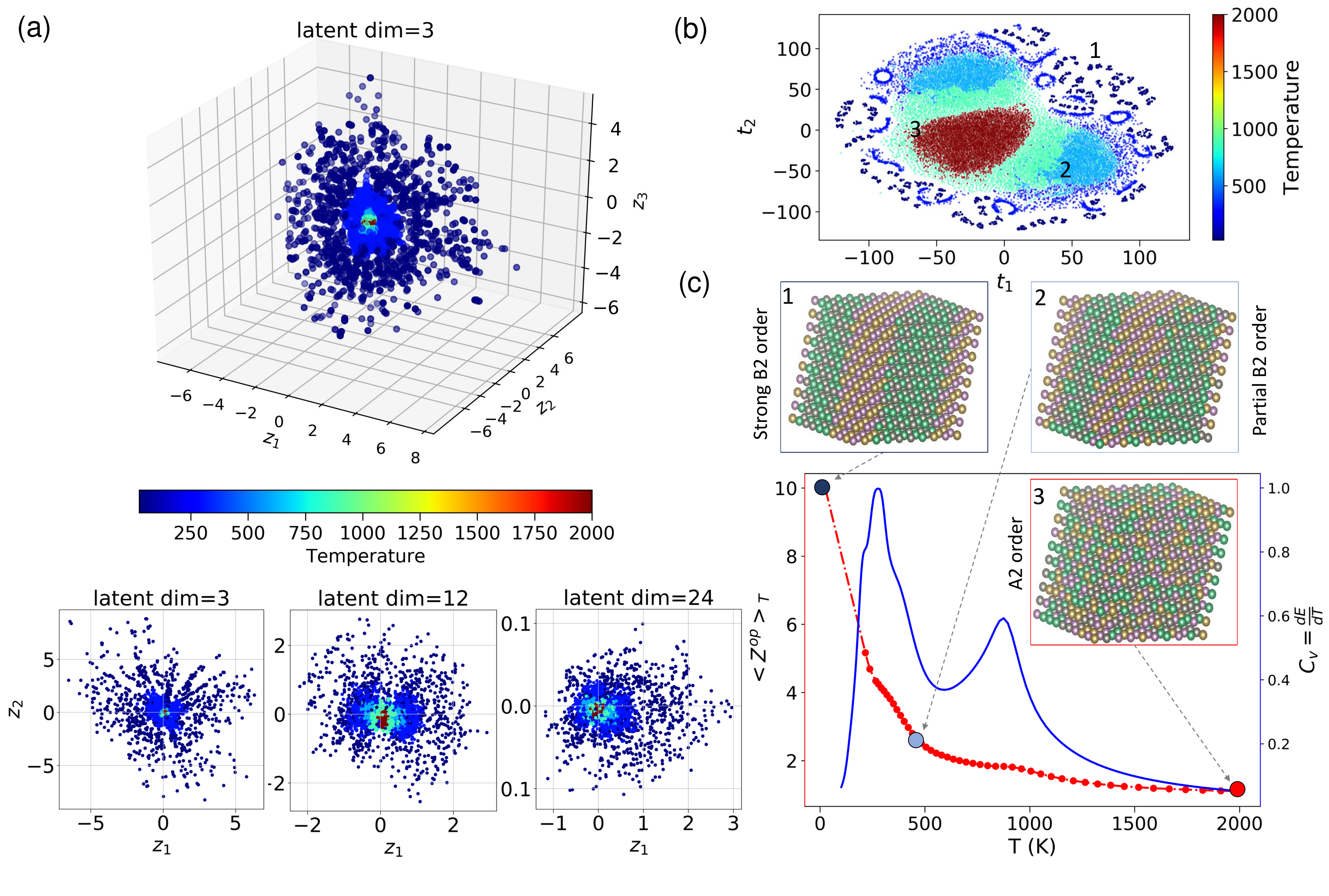}
  \caption{VAE analyses on MoNbTaW. (a) The embedding plots. (top) 3D plot of the VAE model with latent dimension size 3; (bottom) first 2 dimensions ($z_1$ and $z_2$) of the latent space of the VAE model with latent dimension size 3 (left), 12 (middle), 24 (right), respectively. (b) The 2D t-SNE plot of VAE 12D embedding on the test data. (c) The VAE based order parameter $Z^{op}$ (red) and the specific heat $C_v$ (second-order moment of energy) at different temperatures. The inserted snapshots are sample system configurations at 3 different temperatures{\color{black}, showing the microstructural changes from  (1) strong B2 order, (2) partial B2 order, to (3) A2 (BCC) as the temperature increases. The corresponding mapped data points in the latent space are also marked in (b)}. Here a size of $N=10\times10\times10$ is adopted, a different and larger size does not affect our results. }
  \label{fig:tsne4}
\end{figure*}

{\bf Applications to equiatomic HEAs}

In the following experiments, we demonstrate the VAE based order parameter by two real HEAs, i.e., MoNbTaW and MoNbTaVW, and compare with the state-of-the-art Warren-Cowley order parameters. We also examine other properties of an order parameter, such as the size effect and its second-order moment, to further confirm the validity of our approach. 

{\it MoNbTaW ~}
MoNbTaW alloy is one of the important HEAs and has been intensively studied \cite{MoNbTaW-MRL,MoNbTaW-npjcm,MC-hea}. We train our VAE models (see Fig.~\ref{fig:nn-arch}) on MC generated MoNbTaW configurations, setting latent variable $z$ of various size of dimensions. 
The training data is pooled at five different temperatures in the range of $25\sim3,000$ K, and $<Z^{op}>_T$ is then calculated at various $T$s on test data by inference using the trained VAE models. All the reported results are on {\it test data} if not explicitly specified.  In Fig.~\ref{fig:loss}, the validation loss is plotted against the size of latent dimension, each data point corresponding to a VAE model. With higher $z$ dimensions, the model compresses less and hence is more capable. On the other hand, if $z$ dimension is too high, the model is forced to optimize on local features and output can be incoherent. For our model on this data, the validation loss plateaus near the dimension size 12. In terms of the reconstruction accuracy, all models with different $z$ can perfectly predict the BCC sites out of the simple cubic box, which validates that our data representation approach captures the lattice symmetry. If only BCC sites are considered, for samples drawn from $T=25$ K, the more stringent reconstruction accuracy is $\sim 98\%$. At $T=2,000$ K where atoms nearly randomly occupy the BCC sites, the model gives equal probability of $0.25$ for each channel (atom type), therefore the accuracy is $\sim 25\%$.  

The latent spaces are show in Fig.~\ref{fig:tsne4}a for the test data embedding for models with latent dimension size 3, 12, and 24. It is consistent across different dimensions that $z$ for high temperature samples are indeed near the origin in the latent space, while the low temperature ones are scattered around. In the full latent space plot (latent dim$=3$), there are spike lines radiated out from the center, indicating transition ``paths" from ordered states to disordered states. At higher latent dimensions, only first 2 dimensions are plotted, and the data points seem to form two symmetric clusters during the middle range temperatures (more clearly in the case of latent dim$=$ 12). They should correspond to the ordered B2 structures \cite{MoNbTaW-npjcm}, where Mo-W pair and Nb-Ta pair bond separately (see channel breakdown in Fig.~\ref{fig:nn-arch}).            

To better illustrate the clustering of different phases, the 2D t-SNE of the embedding in 12D latent space is plotted in Fig.~\ref{fig:tsne4}b. Data points seem to cluster by temperatures, with a single cluster (disordered phase, more symmetric) at high temperature and many small local clusters at low temperature (ordered phase, less symmetric). Around $T=1,000$ K, the single cluster starts to separate, indicating a phase transition. Two representative configurations (See Fig.~\ref{fig:tsne4-ab} A and B) for the two separated phases correspond to two alternative B2 structures, i.e. more ordered Mo and W versus more ordered Nb and Ta. This explains the symmetry of the plot along $t_1 = t_2$ and also confirms that the symmetry in input configurations is preserved in the latent space. Note that the distance metric in $t$ does not have the same meaning as in $z$ space but the key features are also persevered in t-SNE and intentionally magnified.     

Quantitatively, the order in MoNbTaW system can be described by the VAE based order parameter. Using the trained model to inference on test data, the ensemble averaged $\langle Z^{op}\rangle_T$ is calculated at each temperature following Eq.~\ref{eq:z}. To compare, the Warren-Cowley SRO parameters for the nearest neighbors are also estimated for each pair of atoms, and the Pearson correlation coefficient between $\langle Z^{op}\rangle_T$ and SRO are listed in Table~\ref{tab:pearson}. The distance metric is quite robust. Both L1 and L2 norms behave similarly and strongly correlate with SRO for a range of models with different dimension sizes of the latent space. Overall, the 12D latent variable works best, agreeing with the validation loss across various latent dimensions (see Fig.~\ref{fig:loss}).

The $\langle Z^{op}\rangle_T$ across the full temperature range is plotted in Fig.~\ref{fig:tsne4}c. To show the region of interest, second-order moment of energy ($C_v$) and 3 representative configurations are also plotted. The phase transition point seems to be around 1,000 K, consistent with the cluster separation temperature as shown in Fig.~\ref{fig:tsne4}b. {\color{black} Identifying the type of order is challenging when only order parameters are available, but this challenge can be overcome assisted with the real-space microstructure or other experimental information \cite{feng2021high}.}

\begin{figure*}[!ht]
\centering
  \includegraphics[width=\linewidth]{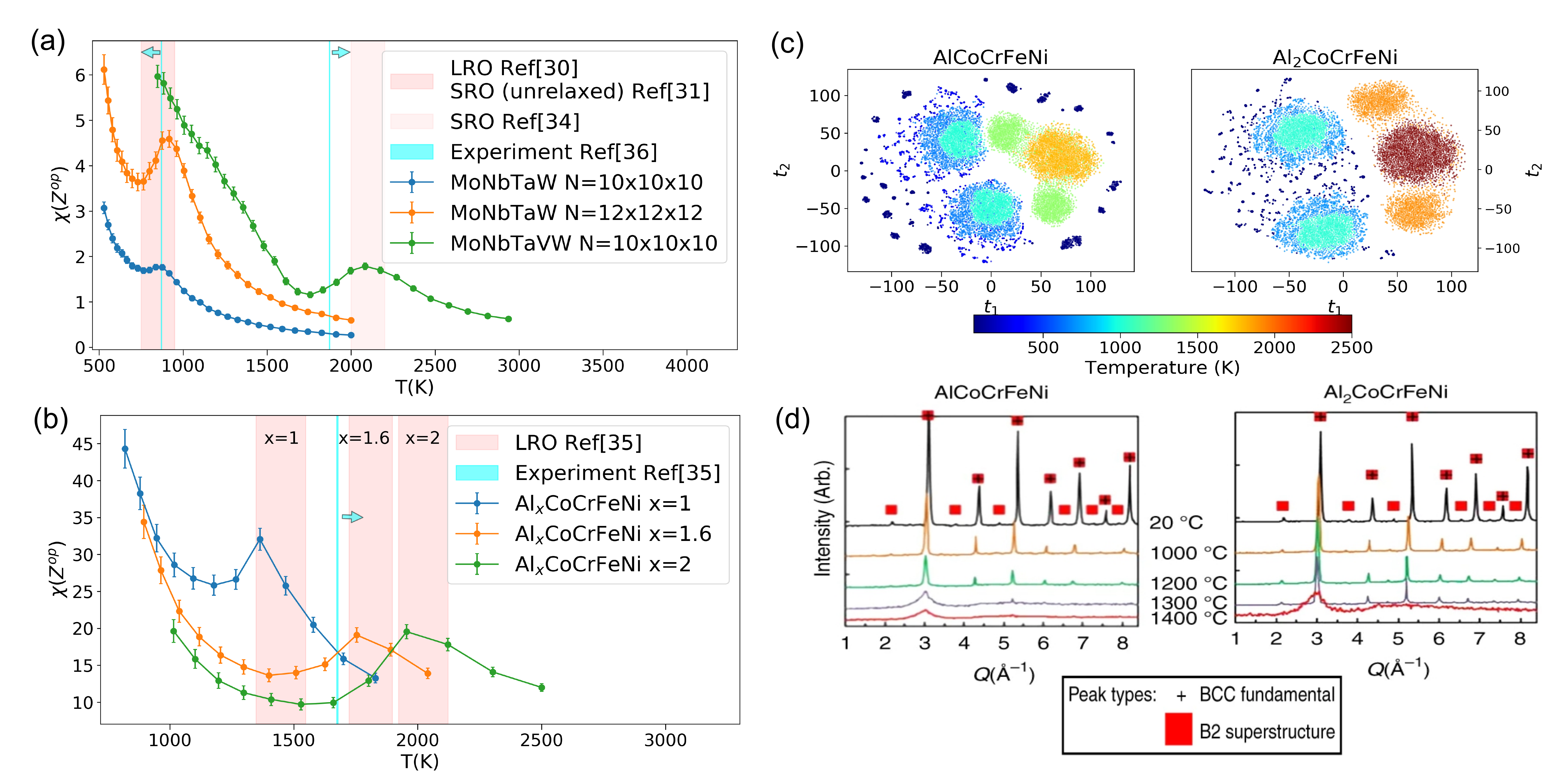}
  \caption{Comparison of our VAE order parameter with experiment. The performance of VAE order parameter is demonstrated along with SRO and LRO parameters (estimated transition temperatures with error ranges are marked by peak bars), and experimental data (transition temperatures marked by cyan bars). The second-order moment ($\chi(Z^{op})$) of VAE order parameter is calculated for (a) MoNbTaW of size $10^3$ and $12^3$, and MoNbTaVW of size $10^3$. (b) Al$_x$CoCrFeNi of size $12^3$ for $x=1$, 1.6, and 2. Error bars are standard errors of the means estimated from the data ensemble at each $T$, which consists of 13,000 to 20,000 samples. (c) The t-SNE clustering of configurations of Al$_x$CoCrFeNi in the VAE latent space. (d) Intensity versus scattering vector magnitude (Q) measured by the neutron scattering experiment from Ref\protect\cite{Al-hea}. }  
  \label{fig:x}
\end{figure*}

{\it MoNbTaVW}
We study a five-component HEA, MoNbTaVW, to check if the VAE order parameter can quantitatively evaluate the effect of V element on the phase transition temperature of MoNbTaW, following the same procedure. The VAE based order parameter $Z^{op}$ and second-order moment of energy $C_v$ are shown in Fig.~\ref{fig:cv5}. Both quantities indicate a phase transition at a temperature higher than 1,000K. The ordered configuration seems to follow the same B2 structure as in the four-component system. Additionally, the signal of phase transition can be observed from the clustering of the configurations in the latent space. We apply t-SNE to further reduce the dimensionality into 2D space (See Fig.~\ref{fig:tsne5}). The behavior is similar to the four-component system, and the plot is symmetric along $t_1 = t_2$, which indicates adding element V does not change the symmetry of the alloy in the latent space. The cluster separates from a single cluster at a higher temperature, which is around 2,000K. This is consistent with $\langle Z^{op}\rangle_T$ and $C_v$ curves in Fig.~\ref{fig:cv5}.   

{\bf Applications to non-equiatomic HEAs}

{\color{black}
{\it Al$_x$CoCrFeNi}
Beyond the two equimolar HEAs, we investigate a widely acknowledged five-component systems, Al$_x$CoCrFeNi, to confirm if the VAE order parameter is sensitive to the concentration changes in HEAs. We use the same interaction parameters in a previous study \cite{Al-hea}, and apply the VAE approach to study the phase transition with the concentration of Al ranging from $x=1$ to $2$. Following the t-SNE clustering on configurations in the latent space, as shown in Fig.~\ref{fig:x}c, the two-step phase transformation is clearly indicated. The first cluster separation happens around 1,400K and 2,000K for Al$_1$CoCrFeNi and Al$_2$CoCrFeNi, respectively, which corresponds to B2 phase order-disorder transition. The second cluster separation is around 700K for both systems, signaling the partially ordered B2 to coherent phase mixtures with disordered BCC and strongly ordered B2. This is in excellent agreement with the reported Monte Carlo simulations \cite{Al-hea}.         
}

{\color{black}
To further validate the VAE order parameter, we calculate the second-order moment of $Z^{op}$, i.e. $\chi(Z^{op})$ as defined in Eq.~\ref{eq:x}, the peak location of which is another strong indicator of a phase transition. As shown in Fig.~\ref{fig:x}a, the $\chi(Z^{op})$ for MoNbTaW and MoNbTaVW are plotted (for size $10^3$), with the peak locations around 900K and 2,090K (marked with dashed lines), respectively. To show the size effect, we also calculate the susceptibility for a larger system size ($12^3$). Its peak is more pronounced, which is a common size effect also existent for other quantities, such as the specific heat. The most significant feature is the consistent peak locations across different system sizes, making it a reliably alternative to specific heat to find transition temperatures. Furthermore, similar evidence is shown for Al$_x$CoCrFeNi (see Fig.~\ref{fig:x}b), where the peak of $\chi(Z^{op})$ (indicating order-disorder transition temperature) moves from around 1,400K to 2,000K when Al concentration increases from $x=1$ to $2$. This is consistent with clustering behavior shown in Fig.~\ref{fig:x}c.     
\begin{figure*}[t]
\centering
  \includegraphics[width=\linewidth]{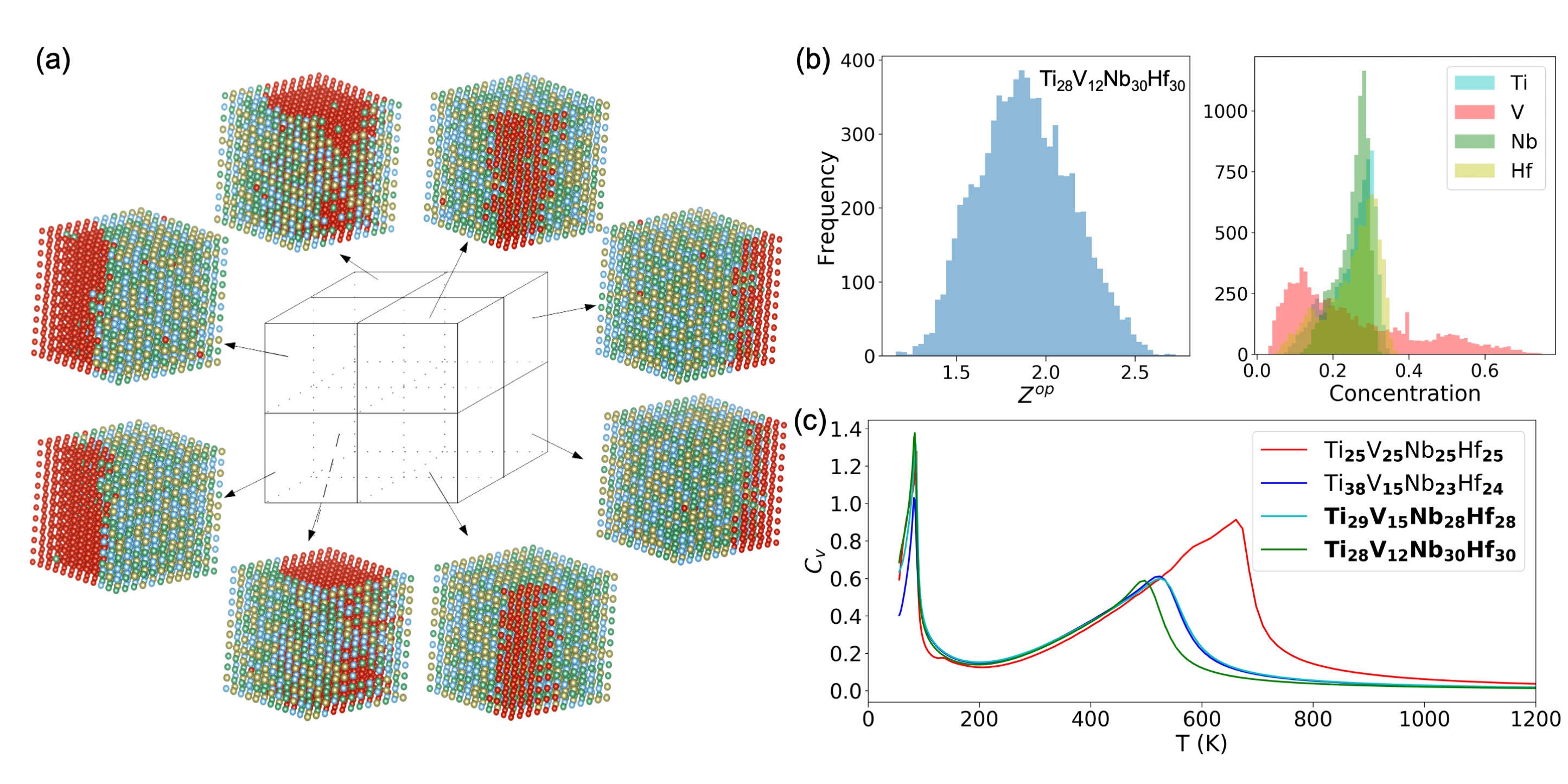}
  \caption{VAE informed procedure for HEA design. (a) A large number (e.g., 1,000) of Monte-Carlo frames at a certain temperature (e.g., 500 K) are taken and each frame is subdivided into 8 subsystems, giving 8,000 samples in the first step. (b) The VAE order parameter and the element concentrations are then calculated for each sample. The most promising systems are identified as the ones with highest frequency in the $Z^{op}$ histogram. (c) Our Monte-Carlo calculations confirm the selected systems indeed push down the ordering transition temperature compared to the equiatomic system, and the method yields systems preferred by thermodynamics, or nature.} 
  \label{fig:design}
\end{figure*}
We compared our VAE order parameter with both the existing order parameter in HEA, i.e. Warren-Cowley SRO and long-range order (LRO) parameters, and the experimental data, in Fig.~\ref{fig:x}a and Fig.~\ref{fig:x}b. For all five HEAs we evaluated, the transition temperatures agree with reported SRO \cite{MoNbTaW-npjcm,MC-hea} and LRO results \cite{MoNbTaW-MRL,Al-hea}, respectively, while the VAE order parameter is more generic, i.e. it does not require to define reference lattice types like LRO, and more scalable, i.e. it provides a single scalar indicator rather than a distance-dependent pairs such as SRO. Comparing with the yield-stress \cite{experiment1} and neutron scattering \cite{Al-hea} experiments, the VAE order parameter is consistent as well. 

The neutron scattering experiment in Fig.~\ref{fig:x}d provides further information on the phase transition of this system. It shows the B2 ordering is stronger with higher Al concentration. For example, the peaks for B2 superlattice for Al$_1$CoCrFeNi disappear above 1,273 K (1,000$^{\circ}$C), while for Al$_2$CoCrFeNi these peaks survive even at 1,573 K (1,300$^{\circ}$C). This is consistent with our VAE results in Fig.~\ref{fig:x}b.
}

{\bf VAE order parameter assisted alloy design}

The previous examples have demonstrated the success of applying the VAE-based order parameter to describe ordering degrees in alloys. As a representative parameter, our VAE order parameter can well represent the ordering state of a system, or the sub-system (sub-block) of a large system. This encourages us to explore its potential to guide alloy design through simulated annealing or simulated nature mixing, following the experimental version of the method \cite{wei2020natural}. Since the atomic interactions between different chemical species are different, equiatomic HEAs can never be thermodynamically the most favorable compared to non-equiatomic systems. After annealing the equiatomic mixture at a certain temperature, the initially homogeneous system becomes heterogeneous, forming various phases. 

It requires a much larger system size of the equiatomic system to realize the simulated nature mixing procedure. Here we adopt a system of 13,824 ($24\times 24 \times 24$) atoms. We use a block of size $12\times 12 \times 12$ to subdivide the whole system, giving 8 sub-systems for each system (Fig. \ref{fig:design}a). The specific heat curve of the equiatomic TiVNbHf shows the annealing temperature can be chosen around 500 K, or any other temperature between the two order transitions (\ref{fig:design}c). After thermodynamic equilibrium at 500 K, we take 1,000 frames, resulting in 8,000 sub-systems. A histogram is plotted to find the systems with the highest frequencies (Fig. \ref{fig:design}b). We use $Z^{op}$ histogram to identify the sub-systems with low V concentrations, since our DFT calculations show V atoms do not like any other chemical species in the system. This yields two systems, Ti$_{28}$V$_{12}$Nb$_{30}$Hf$_{30}$ (dominant) and Ti$_{29}$V$_{15}$Nb$_{28}$Hf$_{28}$. Interestingly, the latter system with the highest frequency has a similar component with Ti$_{38}$V$_{16}$Nb$_{23}$Hf$_{24}$, a system identified experimentally through a thermodynamic method ``nature mixing" \cite{wei2020natural}, in terms of the V concentration. We also find another interesting connection between our $Z^{op}$ parameter and the concentration of vanadium, i.e., lower $Z^{op}$ corresponds to systems of lower V concentrations. The two systems occupying the largest space (the peak of $Z^{op}$ histogram) have low V concentration, consistent with the experimental finding. Since V atoms do not like the other three atoms, its concentrations span the whole concentration range from 0 to approximately 1. The highest frequencies are dominated by the zero end, showing V is rejected from the initially homogeneous system. The estimated $T_c$'s of the systems also confirm it. The low-V systems all have lower $T_c$ than the equiatomic one, delaying the ordering transition from homogeneous system to heterogeneous (Fig. \ref{fig:design}c), which is exactly a property needed for ductile HEAs. 

\begin{figure}[!ht]
  \includegraphics[width=\linewidth]{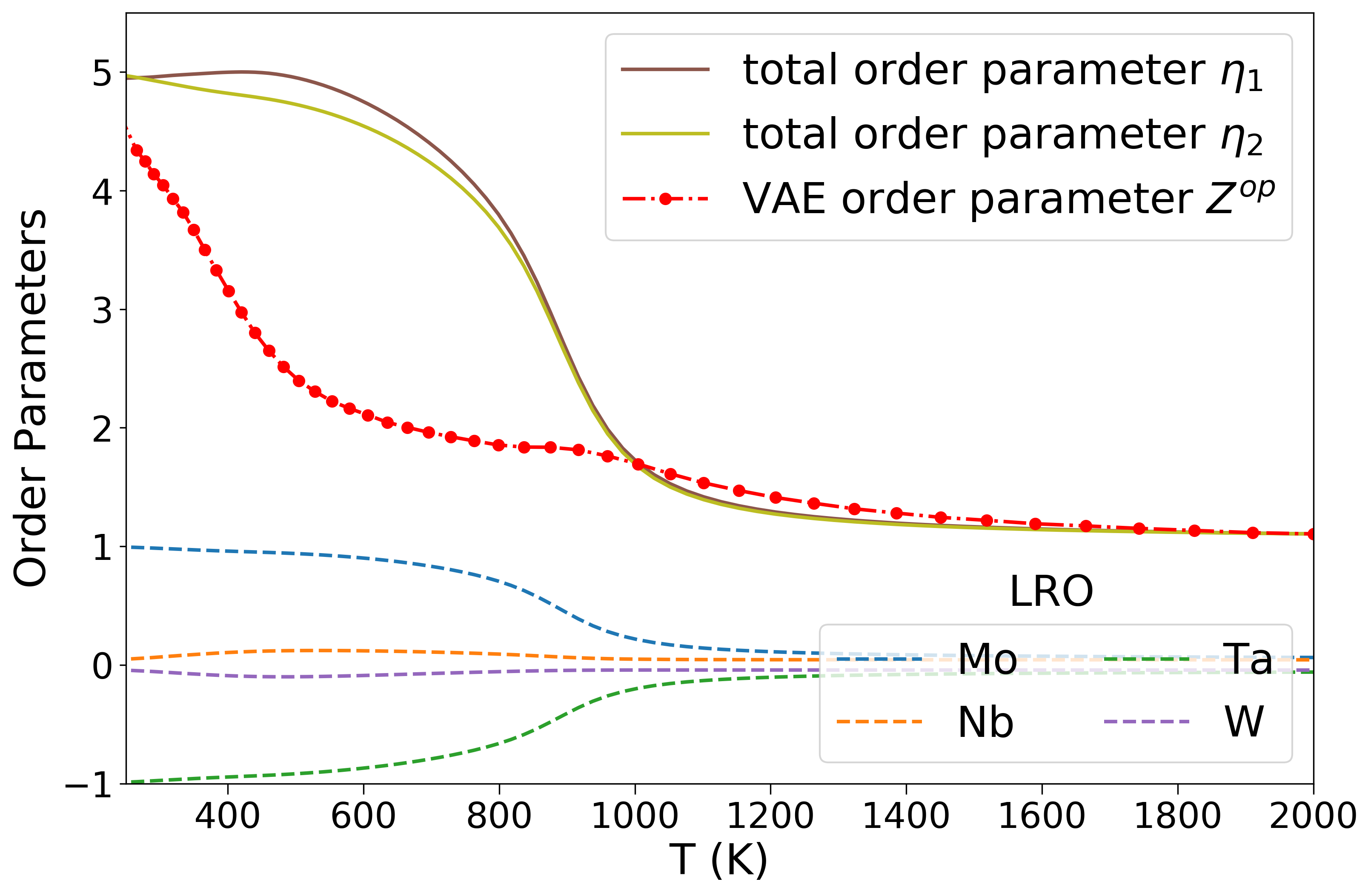}
  \includegraphics[width=\linewidth]{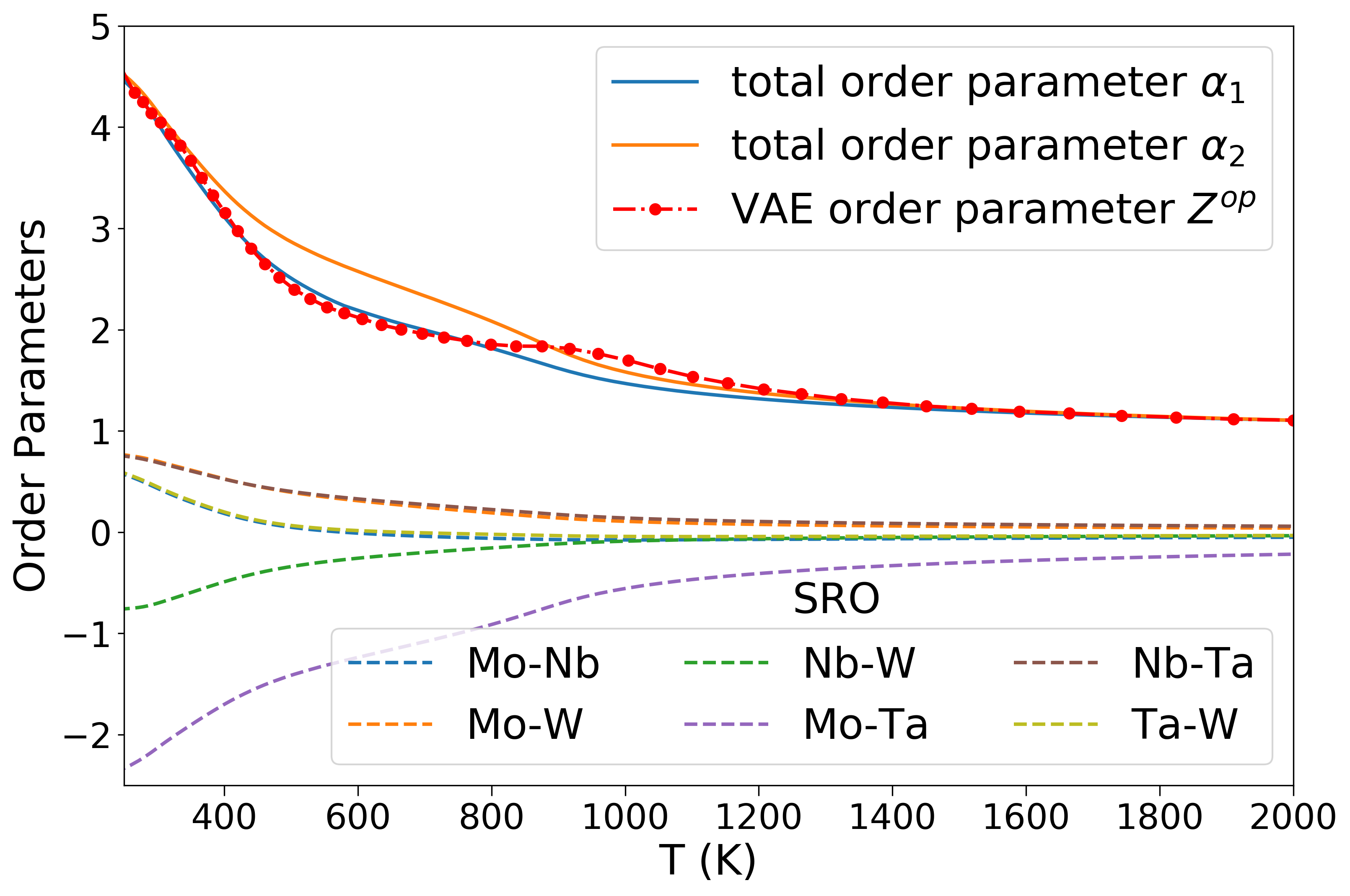}
  \caption{\color{black} Comparison between our VAE order parameter and conventional order parameters. The total LRO fits well with $\langle Z^{op}\rangle_T$ at the high temperature range, while the total SRO fits well with $\langle Z^{op}\rangle_T$ at the low temperature range.
  Either total LRO or SRO parameters only capture a part of the trend of our VAE order parameter. Compared to the total LRO parameters, our VAE order parameter $\langle Z^{op}\rangle_T$ is more similar to the total SRO parameters. This demonstrates $\langle Z^{op}\rangle_T$ captures the total variance of individual Warren-Cowley order parameters (SRO). It is indeed possible to represent the ordering degree of one system using this single number rather than a number of individual order parameters. However, our VAE order parameter can capture the most significant variances of separate ones better than the total order parameters, showing it cannot be obtained by simple operations based on the separate ones. This can only be obtained from the powerful VAE algorithm.}
  \label{fig:total-order}
\end{figure}
The simulated nature mixing, assisted by our VAE order parameter, is shown to be useful for alloy design, which is not limited by the database, since the key properties of all relevant elements for HEAs can be accurately calculated by DFT. After running MC simulations and machine-learning, our new VAE-based order parameter can help identify the good candidates for new HEAs, depending on the targeted properties.




\section*{Discussion}



{\color{black}
{\bf Comparison between our VAE order parameter and the conventional ones}

Our VAE order parameter is calculated based on all atom occupations in the lattices, which does not depend on the choice of atom pairs and their neighboring distance. It is able to better capture the random state of a system than individual order parameter and is more similar to their summation.
Here we directly compare $Z^{op}$ with the conventional order parameters, and show its analogue to the so-called ``total order parameter", i.e., a combination of the separate order parameters. We have different choices to define a total order parameter. Based on the LRO parameters, two natural choices for the total order parameters are
\begin{linenomath}\begin{equation}
\eta_1=c_0\sum_i c_i |\eta_{i}|, ~~~ \eta_2^2=c_0\sum_{i} c_i \eta_{i}^2,
\end{equation}\end{linenomath}
where $\eta_{i}$ is the LRO parameter for chemical species $i$, $c_i$ is its concentration and $c_0$ is a renormalization factor.
Similarly, for SRO parameters, we also have two total order parameters,
\begin{linenomath}\begin{equation}
  \alpha_1=c_0\sum_i c_i c_j |\alpha_{ij}|, ~~~
  \alpha_2^2=c_0\sum_{i} c_i c_j \alpha_{ij}^2,  
\end{equation}\end{linenomath}
where $\alpha_{ij}$ is the Warren-Cowley order parameter for atom pair $i,j$. A similar definition was adopted also in Ref. \cite{Liaw2015}
Since each order parameter $\eta_i$ or $\alpha_{ij}$ is a function of temperature, the total order parameter is also a function of temperature. {\color{black} Taking MoNbWTa as an example, we plot $\eta_1(T), \eta_2(T)$, $\alpha_1(T)$ and $\alpha_2(T)$ in Fig. \ref{fig:total-order}. The individual LRO and SRO parameters are also shown as reference. Overall, the individual order parameters have simpler variance with temperature than the total order parameters and $\langle Z^{op}\rangle_T$. At high temperatures, $\eta_1(T)$, $\eta_2(T)$ and $\langle Z^{op}\rangle_T$ have similar trend and capture the ordering transition around 1,000K. At low temperatures, their behavior becomes diverse. $\eta_1(T)$ shows a small peak around 500K, signal of another phase transition. However, the drawback of this simple definition is clear, i.e., it is dominated by the order parameters with larger variances. Our VAE order parameter shows a steep increase around 500K, and captures the phase transition. The very different trend between VAE order parameter indicates it captures more information than LRO.

Compared to $\eta_1(T)$ and $\eta_2(T)$, both total order parameters $\alpha_1(T)$ and $\alpha_2(T)$ are much more similar to $\langle Z^{op}\rangle_T$ and can well capture the key trend of the order parameter, particularly at the low temperature region. $\alpha_1(T)$ has a surprising well fitting with the VAE order parameter. This comparison directly shows that our VAE order parameter indeed has the physical meaning of ``total order parameter", in contrast to the individual order parameters. Still, the deviation between $\langle Z^{op}\rangle_T$ and $\alpha_1(T)$, $\alpha_2(T)$ at 800-1,200K is clearly seen, which may be due to the missing longer range ordering in the total order parameters. Only the nearest-neighbor ordering is included in $\alpha_1(T)$, $\alpha_2(T)$. Further investigations into the details are needed, and this study initiates and lays the foundation for this direction.
}

}

We need to assume many preconditions to unambiguously define the conventional order parameters. Here, our definition for the new VAE-based order parameter is representative and simple for interpretation and application. For example, it does not need to tell the specific neighbor distances, atom pairs, etc. The information is highly concentrated into one scalar number rather than a matrix like the Warren-Cowley order parameters. It is also independent of the number of elements, making it particularly suitable for complex concentrated alloys with many components. It is analogue to X-ray diffraction or other instruments with similar functions, directly transforming atomistic structures (dataset of atom positions) in real space into a reduced space with few principal variables (latent space for VAE and reciprocal space for X-ray). 

{\color{black} As a computational study based on neural networks, the primary motivation is not to pursue a complete description of the phase transitions, but more about a novel and efficient order parameter for the complex multi-principal element alloys. Unlike physical models \cite{ducastelle2012chemical, Johnson2015, Johnson2018}, our scalar VAE order parameter may not capture all physical information on the phase transitions, but it is sufficient to pinpoint the transition points, similar to the typical order parameter for spin systems (i.e. magnetization for classical Heisenberg model, which is defined in L1 or L2 norm). Furthermore, the number of components used to obtain our VAE order parameter, i.e., the latent-space dimensions are 12D (See Figure~\ref{fig:loss}). The scalar order parameter can be arguably deemed as the first principal component of the order parameters.}


Although defined through neural networks, its physical meaning is transparent and simple, where the crystallographic symmetry during phase transitions is quantitatively preserved. It is surprising to see the physical information can be so precisely preserved after going through the highly nonlinear neural-network operations. {\color{black} This is an excellent example to demonstrates the power of machine learning to understand the physics in materials science. Its power is far beyond its common applications as a mathematical black box. More future studies are merited to fully explore its power. }

The quantitative power of our VAE order parameter to describe the important physical properties, either to describe ordering degree or transition temperatures, is fully explored. This is the first kind of its applications to unravel physics assisted by machine-learning methods, not only qualitatively but also quantitatively. More interesting, we show its application not only to describe existing alloys, either equiatomic or non-equiatomic; but more importantly, the neural-network based order parameter can be used to explore for new alloys. Aided by simulated annealing, we show the order parameter is very useful to identify the most profuse systems during the nature mixing process with lower transition temperatures. 


{\color{black}
\begin{figure}[!ht]
  \includegraphics[width=\linewidth]{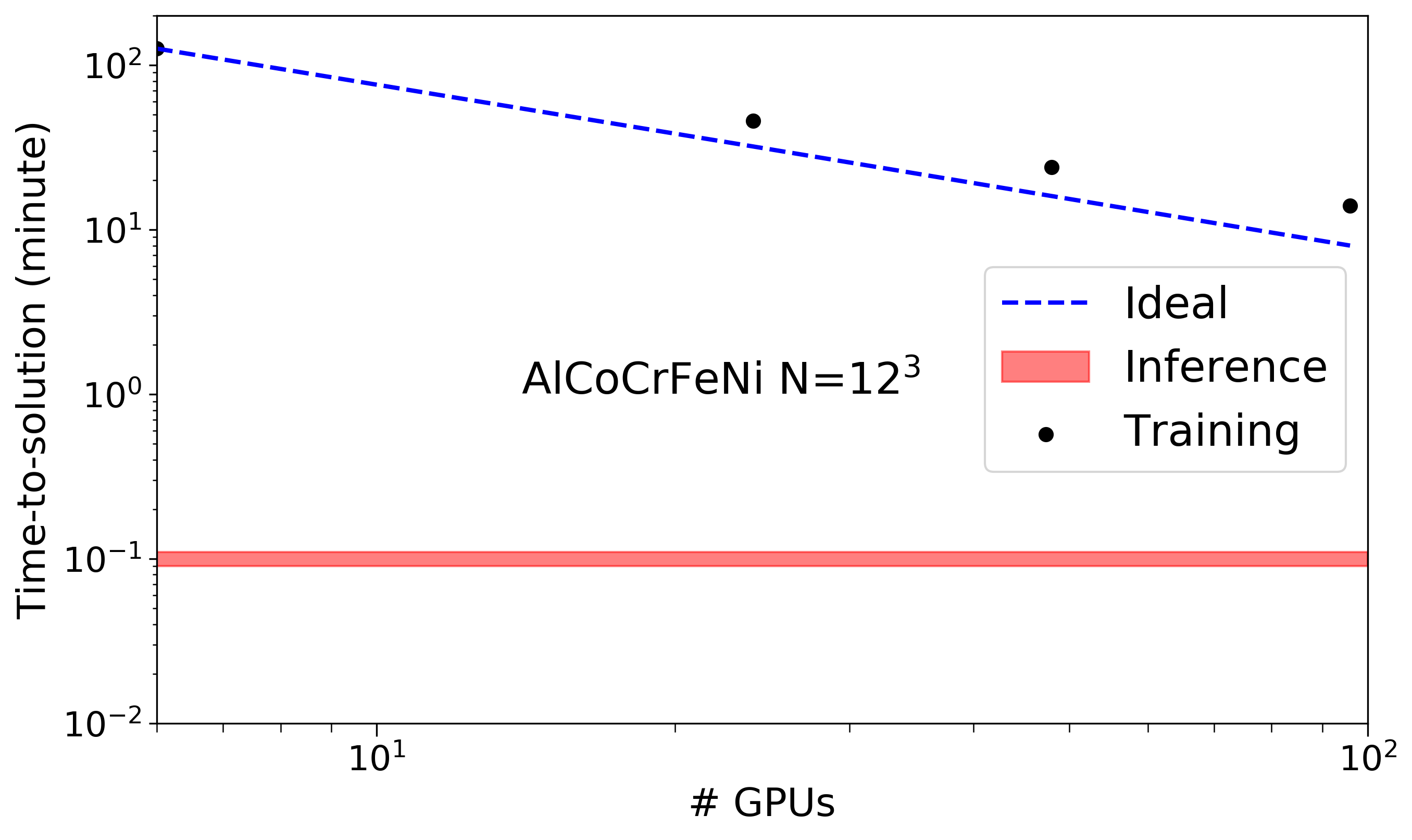}
  \caption{{\color{black}The time to solution for training and inference of AlCoCrFeNi of size $12^3$. The distributed training shows near ideal strong scaling up to 96 GPUs, and the inference latency per GPU is about 0.1 second.}}
  \label{fig:scale}
\end{figure}

{\bf Scaling}

The computational performance of the VAE-based order-parameter approach is shown in Fig.~\ref{fig:scale}. The strong scaling is computed for the VAE training time up to 96 GPUs. The time-to-solution scales almost linearly and the training of a five-component HEA of $12^3$ atoms can be achieved in around 10 minutes. Once the model is trained, it can be used to evaluate the VAE order parameter $\langle Z^{op}\rangle_T$ of a configuration at any temperature in about 0.1 second.}

In summary, we propose a VAE-based order parameter to study the order-disorder phase transition, and demonstrate the approach by establishing the first baseline for physically interpretable VAE on real complex concentrated alloys. The phase transitions are not only indicated qualitatively via the clustering analyses, but also calculated quantitatively by the Manhattan distance metrics in the latent space for the first time. The VAE model provides a generic way to measure the ordering degree in a system, and its second-order moment accurately predicts the ordering transition temperatures. It is surprising to see the physical information can be so precisely preserved after going through the highly nonlinear neural-network operations, demonstrating the huge, unexplored potential of machine learning to understand physics, which merits further explorations.
The method is analogue to X-ray diffraction or other instruments with similar functions, directly transforming atomistic structures in real space into a reduced space with few principal variables.
The neural network based order parameter is also shown to be useful in various systems, as well as in new alloys design. {\color{black}As long as the phase features (e.g., microstructures as shown in Figure~\ref{fig:tsne4}(c)) extracted by the encoder are different (described by Eq.~\ref{eq:reg} and \ref{eq:z}), our approach can be a useful tool to identify the phase transition points.} Our concept of ``VAE order parameter" opens new opportunities in the boundary between artificial intelligence and materials science.

\section*{Methods}
\subsection*{Density functional theory calculations}
Density functional theory (DFT) \cite{Hohenberg1964,Kohn1965} simulations are carried out using Vienna Ab-initio Simulation Package (VASP) \cite{Kresse-vasp} to obtain the total energies for the enthalpy of formation. 
The generalized gradient approximation (GGA) parametrized by Perdew-Burke-Ernzerhof (PBE) \cite{Perdew1996} is used to calculate the electronic exchange-correlation interaction, and the Kohn-Sham equation is solved using projector augmented wave (PAW) method \cite{PAW-Bloechl1994}, where the Brillouin zone is sampled using Monkhorst-Pack scheme \cite{Monkhorst-Pack1976}.
The relaxation stops when the energy difference between ionic steps is smaller than 10$^{-4}$ eV.
A plane wave cutoff of 350 eV and the k-point meshes of $8\times 8\times 8$ for Brillouin zone are used. A supercell size of 2 atoms used for pure element and B2 structure in this study. It includes 8 atom bonds between two chemical species in B2, and the interaction parameters are taken as the enthalpy of formation divided by 8. This offers the input for the Monte-Carlo simulations.

\subsection*{Monte-Carlo simulations}
We consider several real HEAs, i.e. MoNbTaW, MoNbTaVW and Ti-V-Nb-Hf alloys. Each type of atoms has equal or non-equal molar concentration on a body-centered cubic (BCC) lattice. The probability of a $N$ atoms configuration $X = \{x_i\}_N$ at temperature $T$ follows the Boltzmann's distribution $\exp(-E(X)/k_BT) $ where $E$ is the total energy of the configuration and $k_B$ is the Boltzmann constant. 

Given $E$ modelled \cite{hea-model} from high-fidelity quantum-mechanical calculations, the input data at various $T$s are then generated via replica exchange Monte Carlo (MC) simulations \cite{paralleltempering} with following transition probability between replica $m$ and $n$, 
\begin{linenomath}\begin{equation}
 W(\{X_m,T_m\}|\{X_n,T_n\})= \mathrm{min}\left[1,\exp(-\triangle)\right],
\end{equation}\end{linenomath}
where
\begin{linenomath}\begin{equation}
\triangle=(1/k_BT_n-1/k_BT_m)(E(X_m)-E(X_n))
\end{equation}\end{linenomath}
Within each replica, an atom $i$ is exchanged with its neighbor $j$ according to classical Metropolis sampling with the acceptance probability  $P_{ij}\propto\mathrm{min}\left[1,\exp(-(E(\{x_i, x_j\})-E(\{x_j, x_i\}))/k_BT)\right]$. Therefore, each MC step consists of an atom pair exchange trail for each atom with its neighbors and a replica exchange trail among different $T$s.   

{\color{black} For each $T$}, $10^7$ MC steps are performed in addition to initial $10^6$ steps discarded as warm-up, and the input configurations are collected from every 100 steps to minimize the effect of autocorrelation between samples. {\color{black} Configurations from 5 or 6 $T$s across the entire temperature range are then pooled together and $10^5$ samples are randomly selected.} The total $10^5$ input data are then randomly shuffled, and split $80\%$ to $20\%$ for training and testing, respectively. {\color{black} The model inference is performed on all $T$s (192 in total), and the error bars are estimated from the standard error of $10^5$ samples at each T.}         

\subsection*{VAE model}\label{model}

Following the classic VAE\cite{vae-tutorial} {\color{black} in the evidence lower bound (ELBO) formulation}, the loss of our model consists of 2 terms , 
\begin{linenomath}\begin{equation}
\mathcal{L} = D_{KL}[\mathcal{N}(\mu, \sigma), \mathcal{N}(0, 1)] + \parallel X-d(z)\parallel^2, \label{loss}
\end{equation}\end{linenomath}
where $D_{KL}$ is Kullback-Leibler divergence between the distribution of the latent variable $z$ ($z=\mu +\sigma \odot \epsilon$ with $\epsilon \sim \mathcal{N}(0,1)$) and a Gaussian distribution, and the second term is the reconstruction loss (expressed in mean squared error and equivalent to the cross entropy \cite{dl-book} for corresponding distributions) between input $X$ and decoded output $d(z)$.

To simplify the discussion, we assume the dimensions in $z$ are orthogonal and independent, and then the $D_{KL}$ term can be written as,
\begin{linenomath}\begin{equation}
    D_{KL} = -\frac{1}{2}(1 + \log \sigma^2 - \mu^2 - \sigma^2), \label{eq:kl}
\end{equation}\end{linenomath}
where $\mu$ and $\sigma$ are the mean and variance of the $z$ vector. 

For HEA input, the multiple atom types can be one-hot encoded into a vector. Then the activation function of decoder's last convolutional layer should be \textit{softmax}, and the reconstruction loss term will be the categorical cross entropy. Another way is to put each atom type into a channel, and use binary cross entropy on flattened input and output. We find both methods produce similar results. 

The model output is the probability at each lattice site, the sum of which should be close to the total number of atoms $N$. The atom type is determined by taking $\textit{argmax}$ on the channel dimension.     
\subsection*{Model implementation}

Our implementation is based on Keras and TensorFlow, and the Horovod library is utilized for data parallelism across multiple nodes. The experiments are performed on the Summit supercomputer \cite{summit}, where each node is equipped with 6 Nvidia V100 GPUs, 2 IBM Power9 CPUs, and 512GB RAM. The average training time per model (a system or a data split or a latent dimension size) is about 1 hours using 4 nodes.

\section*{Data availability}
The sample training and validation data{\color{black}, along with a pre-trained model} for Ti$_{38}$V$_{16}$Nb$_{23}$Hf$_{24}$ are available at  \href{https://doi.org/10.6084/m9.figshare.14417225.v1}{https://doi.org/10.6084/m9.figshare.14417225.v3}. The data for all figures are available at \href{https://code.ornl.gov/jqyin/deepthermo/-/tree/master/data}{https://code.ornl.gov/jqyin\\/deepthermo/-/tree/master/data}.  

\section*{Code availability}
The VAE model training and order parameter inferencing codes are available at \href{https://code.ornl.gov/jqyin/deepthermo}{https://code.ornl.gov/jqyin/deepthermo}. The figures are plotted with the notebook at \href{https://code.ornl.gov/jqyin/deepthermo/-/blob/master/utils/hea\_vae\_analysis.ipynb}{https://code.ornl\\.gov/jqyin/deepthermo/-/blob/master/utils/hea\_vae\_analysis.ipy\\nb}.

\bibliography{ref}

\section*{Acknowledgments}
This research was sponsored by and used resources of the Oak Ridge Leadership Computing Facility (OLCF), which is a DOE Office of Science User Facility at the Oak Ridge National Laboratory supported by the U.S. Department of Energy under Contract No. DE-AC05-00OR22725. This work was also performed in support of the US Department of Energy’s Fossil Energy Crosscutting Technology Research Program, and in part by an appointment to the U.S. Department of Energy (DOE) Postgraduate Research Program at the National Energy Technology Laboratory (NETL) administered by the Oak Ridge Institute for Science and Education.

\section*{Author contributions}
The initial project idea was formulated by J.Y; J.Y and Z.P. designed the experiments;J.Y. performed the data generation, VAE training, and order parameter inference;Z.P. performed all DFT calculations, interpreted the total order parameters, proposed the alloy design concept and realized the initial version of the method. J.Y. and Z.P. wrote the manuscript. J.Y., Z.P. and M.G. analyzed the results and finalized the manuscript. 

\section*{Competing interests}
The authors declare no competing interests. 

\section*{Additional information}
Correspondence and request for materials should be addressed to J.Y. and Z.P. 

\renewcommand{\theequation}{S\arabic{equation}}
\renewcommand{\thefigure}{S\arabic{figure}}
\renewcommand{\thesection}{S\arabic{section}}
\renewcommand{\thetable}{S\arabic{table}}
\setcounter{figure}{0}   
\setcounter{table}{0}   
\setcounter{section}{0} 

\clearpage
\section*{Supplementary material}

\noindent {\bf Model Parameters.}
Four 3D convolutional (encoder) and convolutional transpose (decoder) layers, respectively:
\begin{enumerate}
    \item 64 filters with kernel size (3,3,3) and stride (1,1,1) 
    \item 64 filters with kernel size (3,3,3) and stride (2,2,2) 
    \item 64 filters with kernel size (3,3,3) and stride (2,2,2) 
    \item 64 filters with kernel size (3,3,3) and stride (1,1,1) 
\end{enumerate}
Four 3D average Pooling (encoder) and up sampling (decoder) layers (for system size $N> 12^3$) with filter size (2,2,2), respectively. 2 Dense layers of size (64, 64, 64, 64). 

\begin{figure}[!ht]
  \includegraphics[width=\linewidth]{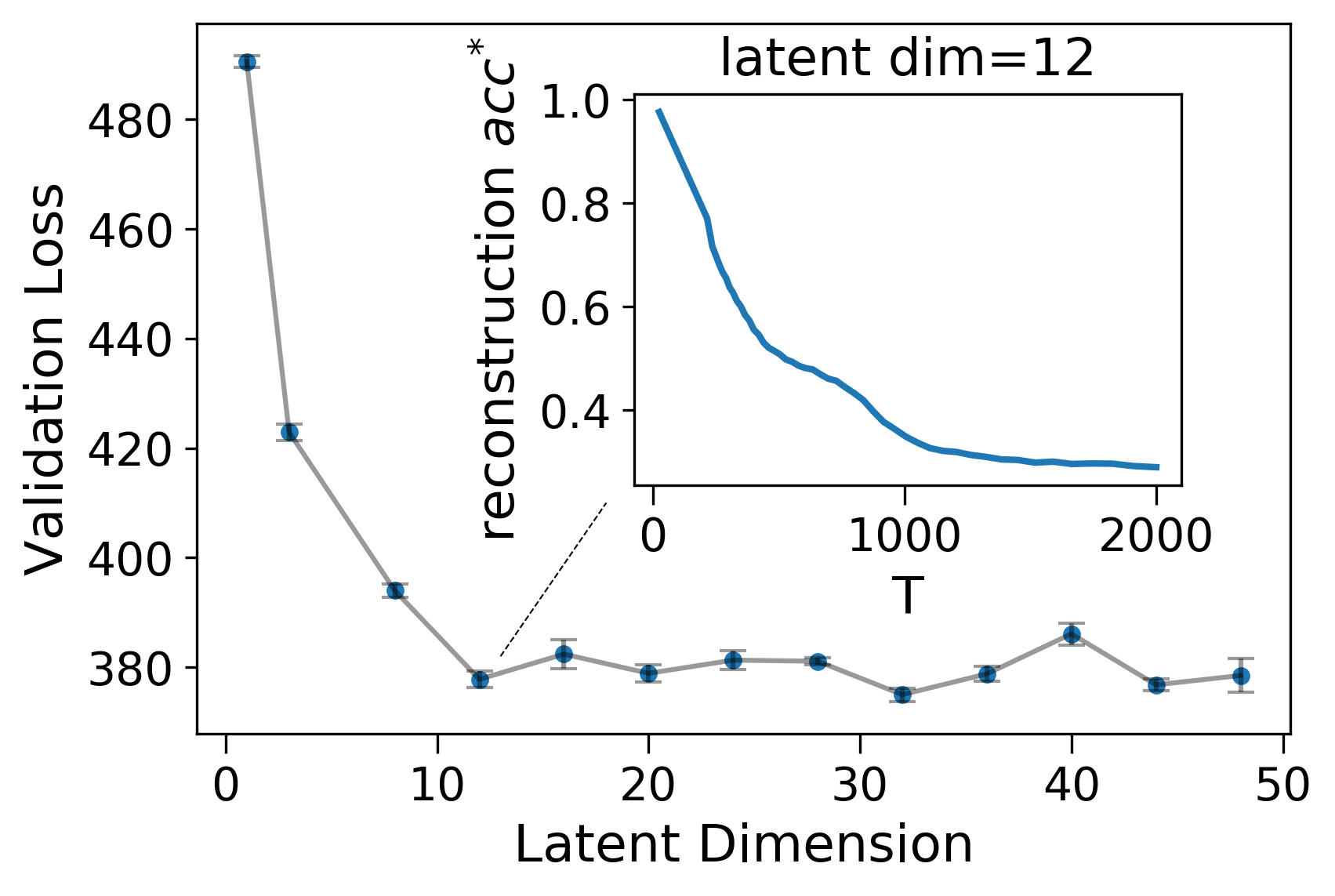}
  \caption{The validation loss versus latent dimension for MoNbTaW of size $N=10\times10\times10$. The insert plot is for reconstruction accuracy$^{*}$ (more stringent since it does not account for the perfect prediction of BCC sites out of a larger simple cubic box) at different temperatures for the VAE model with latent dimension size 12.}
  \label{fig:loss}
\end{figure}

To find the best model, additional hyperparameter search is performed along the latent dimensions, as shown in Figure~\ref{fig:loss}. The validation loss plateaus after latent dimension size 12, the model for which is then selected for the downstream VAE analyses. 

\begin{figure}[!ht]
\centering
  \includegraphics[width=1.05\linewidth]{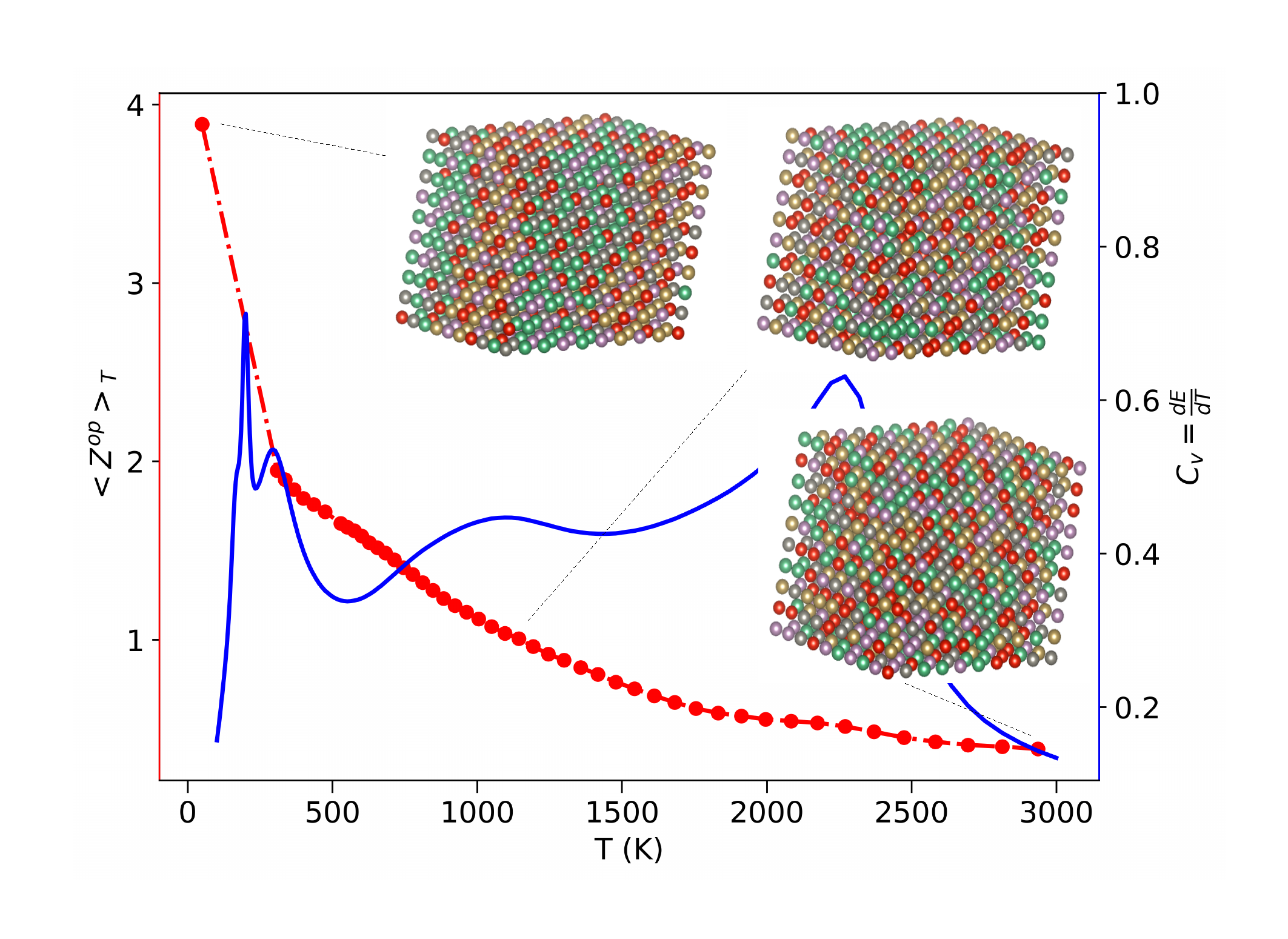}
  \caption{The VAE based order parameter $Z^{op}$ (red) and the specific heat $C_v$ (second-order moment of energy) at different temperatures for MoNbTaVW of size $N=10\times10\times10$. The inserted snapshots are typical system configurations at 3 different temperatures.}
  \label{fig:cv5}
\end{figure}

\begin{figure}[!ht]
  \includegraphics[width=\linewidth]{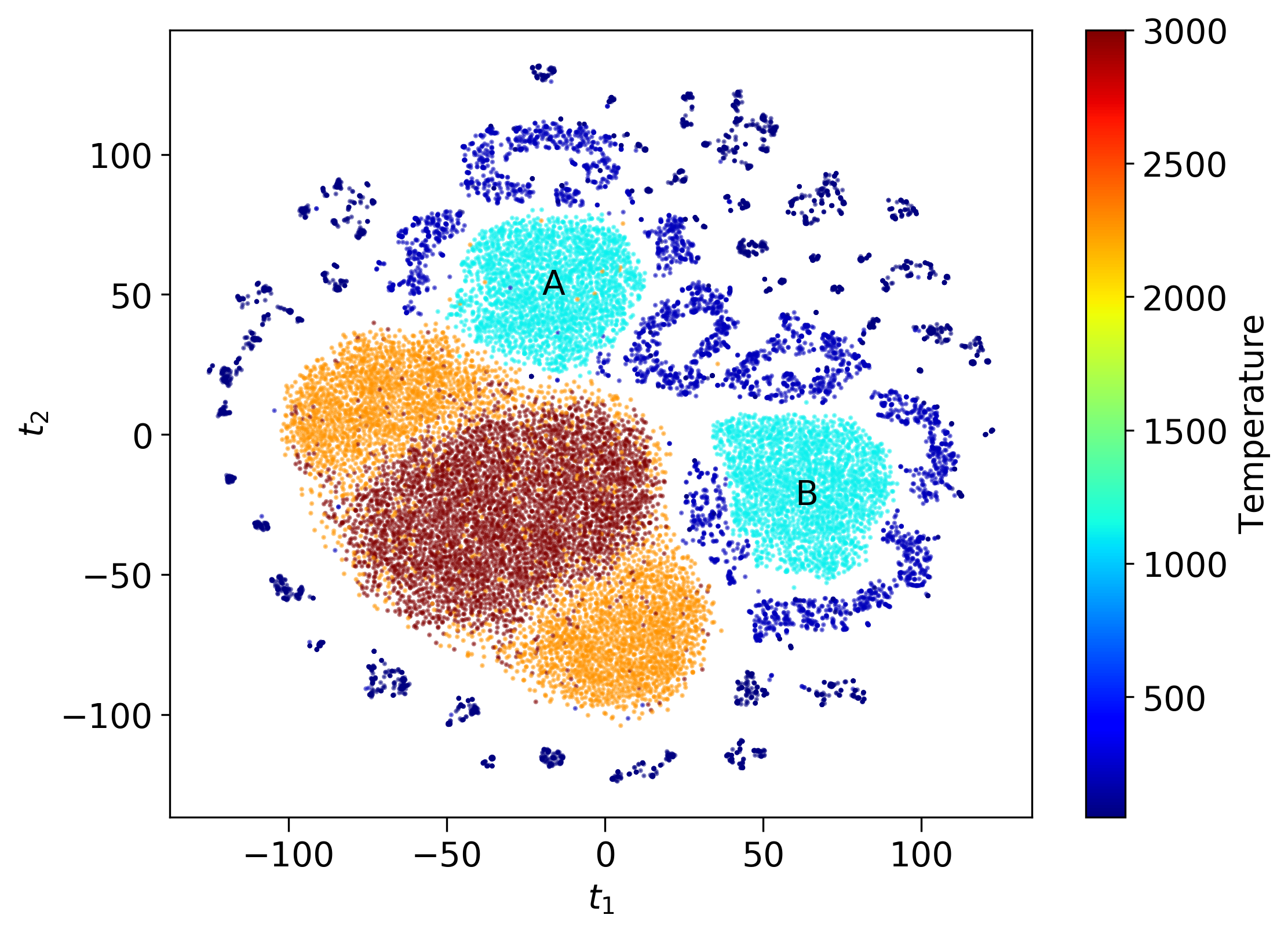}
  \includegraphics[width=\linewidth]{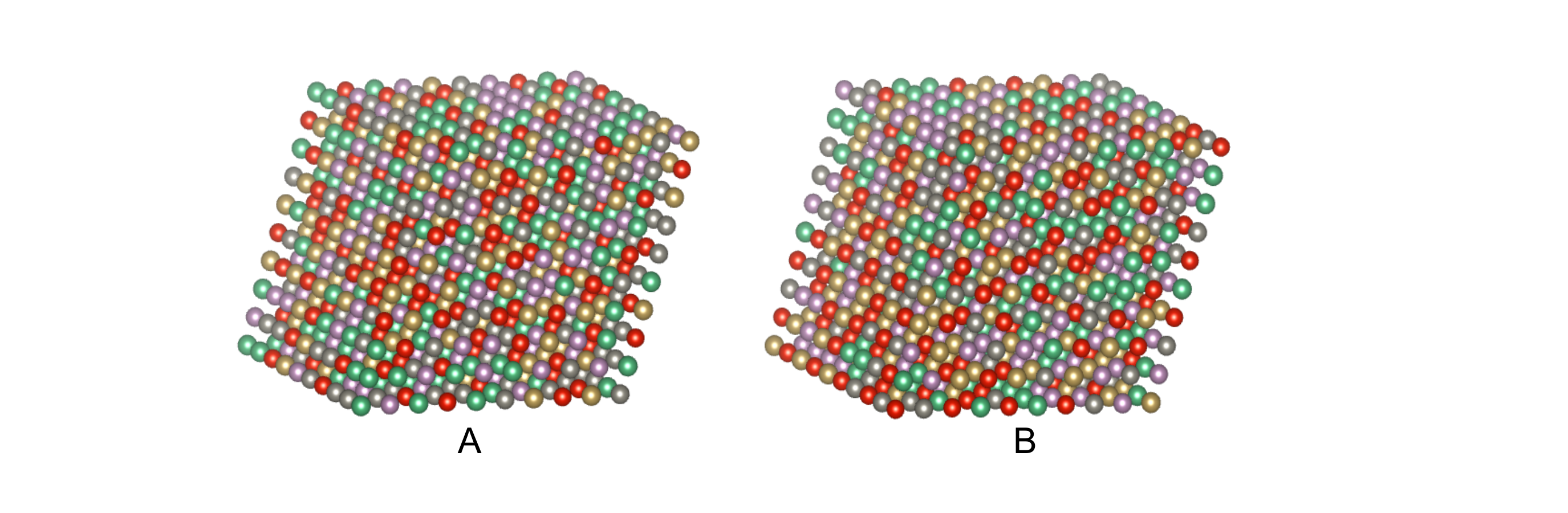}
  \caption{(top) The 2D t-SNE plot of 12D embedding on test data for MoNbTaVW of size $N=10\times10\times10$. (bottom) Two configurations corresponding to the marked (A and B) points in the t-SNE plot.}
  \label{fig:tsne5}
\end{figure}

For MoNbTaVW, the VAE order parameter and t-SNE clustering of the configurations in the latent space are plotted in Figure~\ref{fig:cv5} and Figure~\ref{fig:tsne5}, respectively, and the Pearson correlation coefficient between our VAE order parameters and the Warren-Cowley order parameter are listed in Table~\ref{tab:pearson}. 

\begin{figure}[!ht]
\centering
  \includegraphics[width=1.05\linewidth]{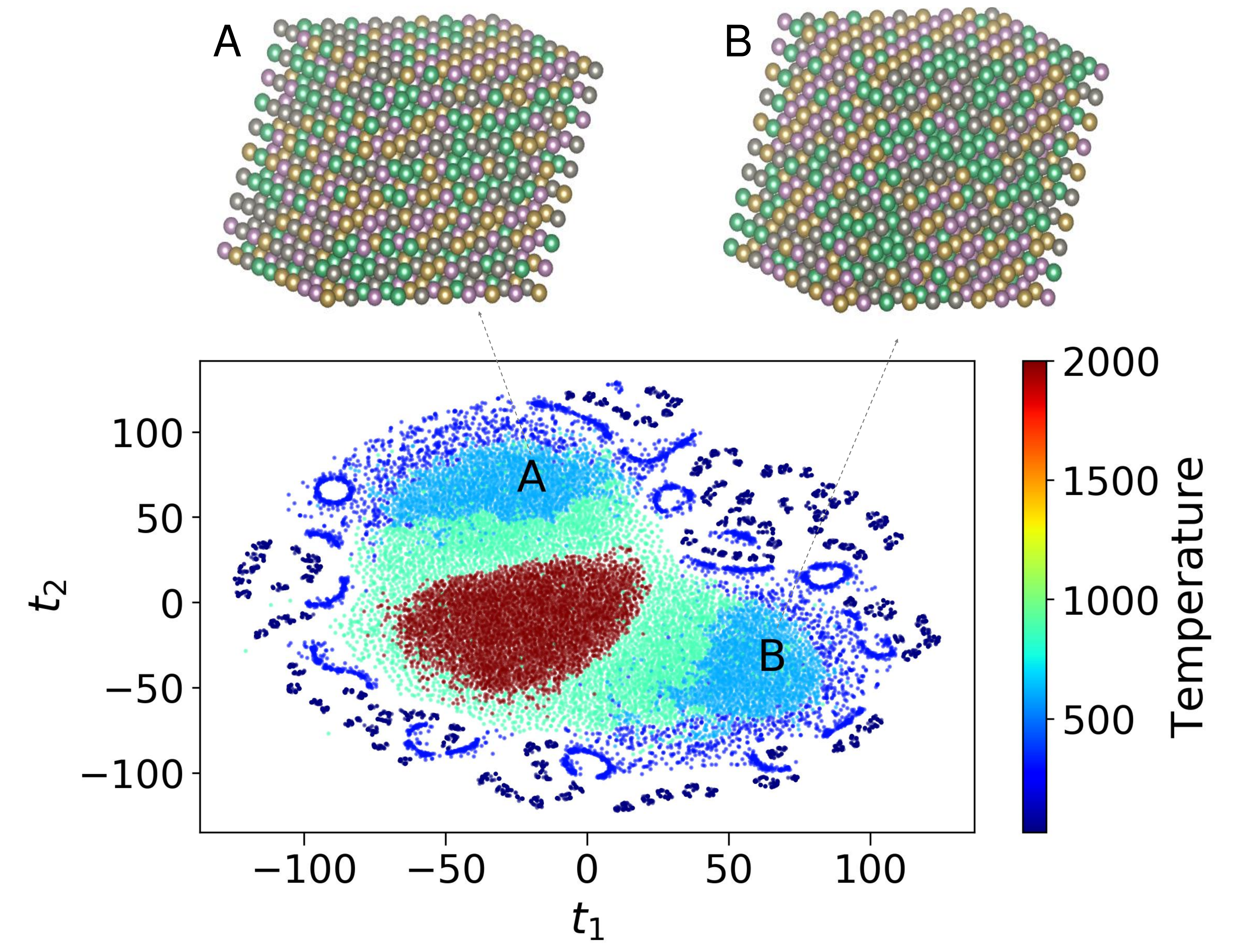}
  \caption{{\color{black}The 2D t-SNE plot of VAE 12D embedding on the test data for MoNbTaW. Two configurations corresponding to the marked (A and B) points in the t-SNE plot.}}
  \label{fig:tsne4-ab}
\end{figure}

\begin{table*}[]
\centering
\begin{tabular}{|c|c|c|c|c|c|c|c|}
\hline
\multicolumn{2}{|c|}{z-distance \textbackslash SRO}                                                   & Mo-Nb          & Mo-Ta          & Mo-W           & Nb-Ta          & Nb-W           & Ta-W           \\ \hline
\multirow{2}{*}{\begin{tabular}[c]{@{}c@{}}latent\\ dim=3\end{tabular}}            & L1 norm          & 0.945          & 0.845          & 0.835          & 0.843          & 0.855          & 0.945          \\ \cline{2-8} 
                                                                                   & L2 norm          & 0.943          & 0.845          & 0.835          & 0.843          & 0.854          & 0.943          \\ \hline
\multirow{2}{*}{\textbf{\begin{tabular}[c]{@{}c@{}}latent \\ dim=12\end{tabular}}} & \textbf{L1 norm} & \textbf{0.959} & \textbf{0.901} & \textbf{0.892} & \textbf{0.898} & \textbf{0.907} & \textbf{0.959} \\ \cline{2-8} 
                                                                                   & \textbf{L2 norm} & \textbf{0.958} & \textbf{0.901} & \textbf{0.891} & \textbf{0.898} & \textbf{0.906} & \textbf{0.958} \\ \hline
\multirow{2}{*}{\begin{tabular}[c]{@{}c@{}}latent\\ dim=24\end{tabular}}           & L1 norm          & 0.940          & 0.853          & 0.844          & 0.852          & 0.864          & 0.940          \\ \cline{2-8} 
                                                                                   & L2 norm          & 0.939          & 0.854          & 0.845          & 0.852          & 0.864          & 0.939          \\ \hline
\end{tabular}
\caption{Correlation between z distance and the Warren-Cowley order parameter. Here we tabulate the Pearson coefficient between z distance metric (L1 and L2 norm) and the Warren-Cowley short-range order parameter (SRO) for MoNbTaW of size $N=10\times10\times10$ across temperature range $25 \sim 2000$ K.}
\label{tab:pearson}
\end{table*}

\begin{figure}[!ht]
\centering
 \begin{picture}(90,420) 
  \put(-70,300){\includegraphics[width=0.86\linewidth]{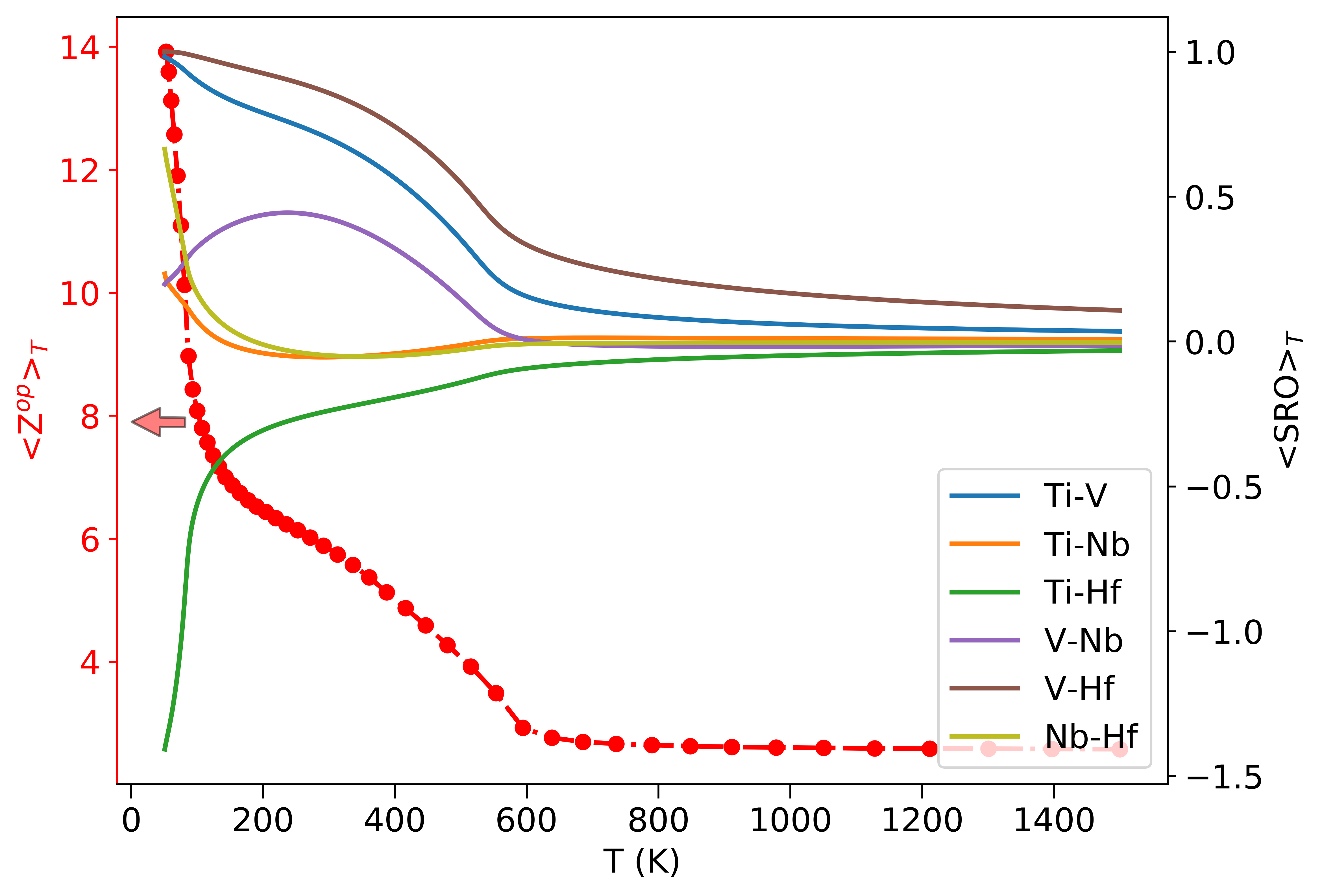}}
  \put(-70,440){{\myfont (a)}}
  \put(-75,140){\includegraphics[width=0.86\linewidth]{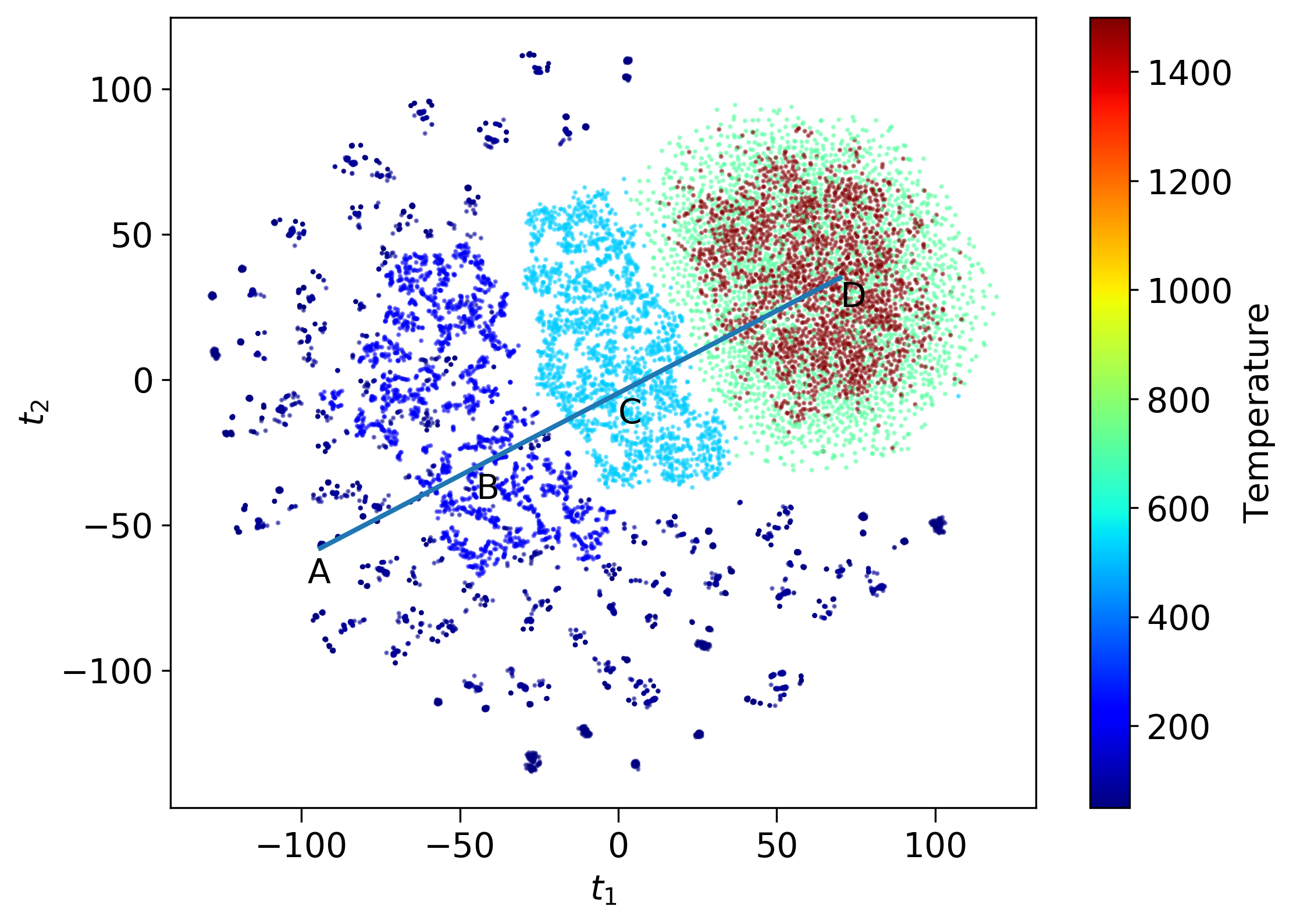}}
  \put(-70,285){{\myfont (b)}}
  \put(-70,0){\includegraphics[width=0.86\linewidth]{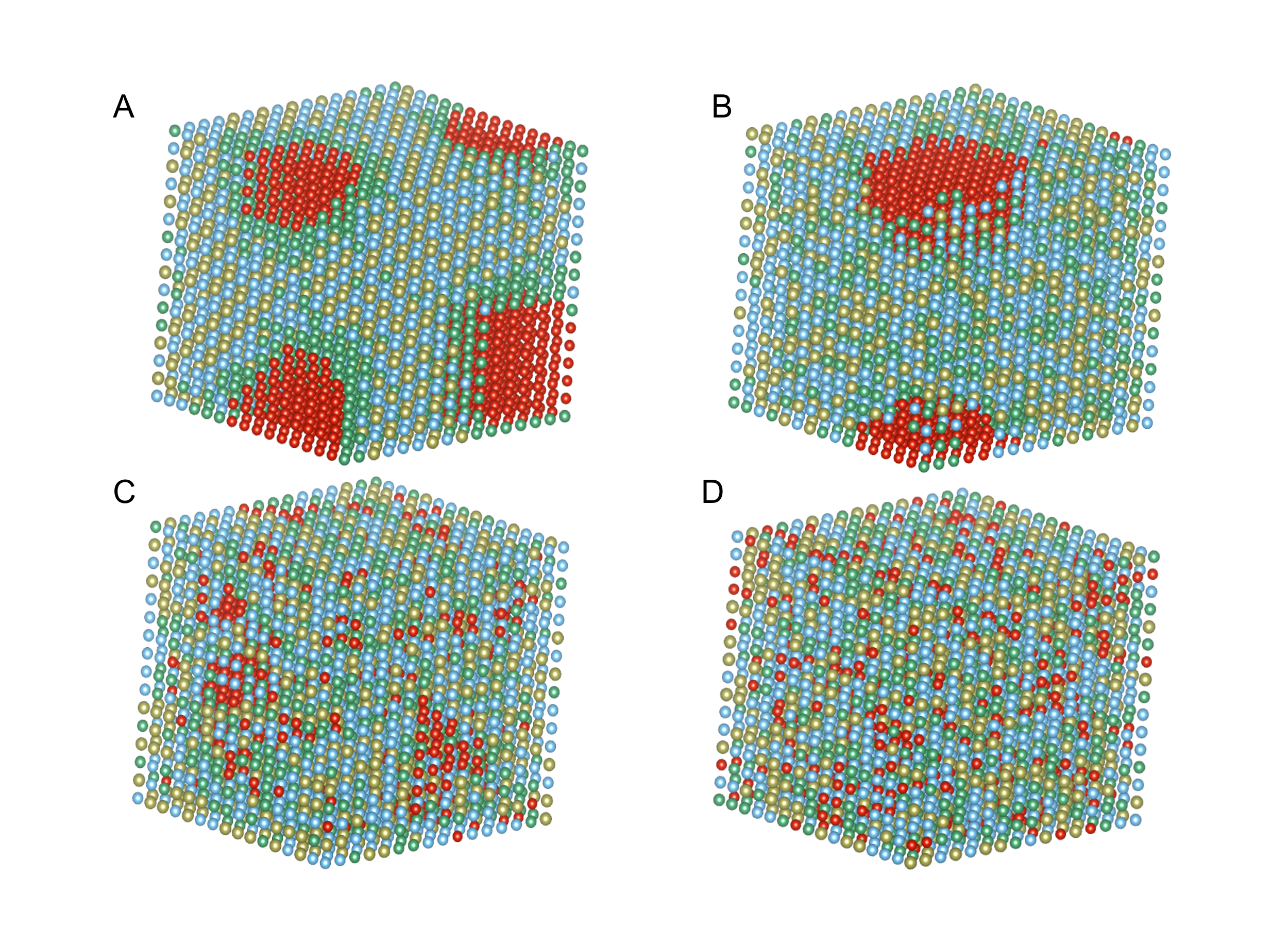}}
  \put(-70,138){{\myfont (c)}}
 \end{picture}
  \caption{VAE analyses on Ti$_{38}$V$_{15}$Nb$_{23}$Hf$_{24}$. (a) The Warren-Cowley short-range order (SRO) (line without marker) and VAE based order parameter $\langle Z^{op}\rangle_T$ is shown for size $N=16\times16\times16$ (line with marker). (b) The 2D t-SNE plot of VAE 12D embedding on test data. (c) Four configurations along the phase transformation pathway marked as A, B, C, and D in the t-SNE plot.} 
  \label{fig:sro_vae}
\end{figure}

\noindent {\bf Ti$_{38}$V$_{15}$Nb$_{23}$Hf$_{24}$}
Here we demonstrate our method on Ti$_{38}$V$_{15}$Nb$_{23}$Hf$_{24}$ system identified experimentally through a thermodynamic method ``nature mixing" \cite{wei2020natural}. The method is basically to (i) anneal equiatomic system for a sufficiently long time and (ii) analyze the available phases formed during the process. The components and their concentrations of the phase with a significantly large area is quantified and used to make a new alloy. Details are referred to the published work.
As is shown in Fig. \ref{fig:sro_vae}a, the $Z^{op}$ order parameter clearly reflects the overall trend of the Warren-Cowley order parameters. Its temperature dependence is dominated by the most significant changes of the Warren-Cowley order parameters, which again confirms it is a representative parameter to describe SRO. We also notice that there is a symmetry along the blue line in Fig. \ref{fig:sro_vae}b, which deviates from the $t_2=t_1$ line appeared in the previous two examples. We already mentioned that the $t_2=t_1$ symmetry is due to two B2-A2 transitions, which are not present here. In this system, the changes of ordering degree are mainly due to the segregation of V atoms from the random matrix. This is confirmed by the four frames at four representative temperatures (red atoms for vanadium in A,B,C,D of Fig. \ref{fig:sro_vae}c). The V atoms are completely random in D, but form small clusters in C, and then become a big cluster in B (still some atoms not in the big cluster) and A.



\end{document}